\documentclass[11pt]{article}
\usepackage{amsmath}
\usepackage{graphicx}
\usepackage{natbib}
\usepackage{url} 
\newcommand{\blind}{0}

\usepackage[linesnumbered, ruled, vlined]{algorithm2e}
\usepackage{booktabs}
\usepackage{natbib}
\usepackage{threeparttable} % add table footnote
\usepackage{amsfonts, amsthm,amssymb}
\usepackage{amsmath}   % add brace above/below the equation \underbrace \overbrace
\usepackage{bm,bbm}    % bbm: identical function
\usepackage{lipsum}
\usepackage{graphicx}
\usepackage{siunitx}
\usepackage{etoolbox}  % adjust the font size of the three part table
\usepackage{adjustbox} % adjust the width of three part table
\usepackage{threeparttable}
\usepackage{adjustbox} % adjust the width of three part table
\usepackage{pbox}

\makeatletter
\newcommand{\algorithmfootnote}[2][\footnotesize]{%
  \let\old@algocf@finish\@algocf@finish% Store algorithm finish macro
  \def\@algocf@finish{\old@algocf@finish% Update finish macro to insert "footnote"
    \leavevmode\rlap{\begin{minipage}{\linewidth}
    #1#2
    \end{minipage}}
  }
}
\makeatother

\SetKwComment{command}{right mark}{left mark} % add comments in the algorithms
\appto\TPTnoteSettings{\footnotesize}  % adjust the font size of the three part table
\graphicspath{{figures/}}  % set the path of figures

\newtheorem{mytheorem}{Theorem}[section]
\newtheorem{mylemma}[mytheorem]{Lemma}
\newtheorem{mydef}[mytheorem]{Definition}

\renewenvironment{abstract}{%
    \small \vspace{-20pt}
    \begin{center}%
        {\bfseries \abstractname \vspace{-.5em}\vspace{0pt}}%
    \end{center}%
    \quotation
    }
{\endquotation\vfil\null}

\newenvironment{articleCategory}{%
    \small \vspace{-20pt}
    \begin{center}%
        {\bfseries Article Category \vspace{-.5em}\vspace{0pt}}%
    \end{center}%
    \quotation
    }
{\endquotation\vfil\null}

\newenvironment{conflictOfInterest}{%
    \small \vspace{-20pt}
    \begin{center}%
        {\bfseries Conflict of Interest \vspace{-.5em}\vspace{0pt}}%
    \end{center}%
    \quotation
    }
{\endquotation\vfil\null}

\newenvironment{visAbstract}{%
    \small \vspace{-20pt}
    \begin{center}%
        {\bfseries Graphical/Visual Abstract and Caption  \vspace{-.5em}\vspace{0pt}}%
    \end{center}%
    \quotation
    }
{\endquotation\vfil\null}

\begin{document}

\def\spacingset#1{\renewcommand{\baselinestretch}%
{#1}\small\normalsize} \spacingset{1}

%%%%%%%%%%%%%%%%%%%%%%%%%%%%%%%%%%%%%%%%%%%%%%%%%%%%%%%%%%%%%%%%%%%%%%%%%%%%%%

\if0\blind
{
  \title{\textbf{A Survey of Numerical Algorithms that can Solve the Lasso Problems}}
  
  \author{Yujie Zhao\textsuperscript{1} 
          (ORCID ID:  0000-0003-2896-4955) 
          and 
          Xiaoming Huo\textsuperscript{2} \\
          \hspace{.2cm}\\
          \small{\textsuperscript{1} Biostatistics and Research Decision Sciences Department, Merck \& Co., Inc}\\
          \small{\textsuperscript{2} The Stewart School of Industrial and System Engineering, Georgia Institute of Technology} }
  \date{}
  \maketitle
} \fi

\if1\blind
{
  \bigskip
  \bigskip
  \bigskip
  \begin{center}
    {\LARGE\bf Title}
\end{center}
  \medskip
} \fi

\begin{articleCategory}
  Advanced Review
\end{articleCategory}

\begin{conflictOfInterest}
  The authors have no conflict of interests.
\end{conflictOfInterest}

\bigskip
\begin{abstract}
In statistics, the least absolute shrinkage and selection operator (Lasso) is a regression method that performs both variable selection and regularization.
There is a lot of literature available, discussing the statistical properties of the regression coefficients estimated by the Lasso method. 
However, there lacks a comprehensive review discussing the algorithms to solve the  optimization problem in Lasso. 
In this review, we summarize five representative algorithms to optimize the objective function in Lasso, including iterative shrinkage threshold algorithm (ISTA), fast iterative shrinkage-thresholding algorithms (FISTA), coordinate gradient descent algorithm (CGDA), smooth L1 algorithm (SLA), and path following algorithm (PFA).
Additionally, we also compare their convergence rate, as well as their potential strengths and weakness.
\end{abstract}

\noindent%
\vfill

\begin{visAbstract}
  \includegraphics[width = 0.65\textwidth]{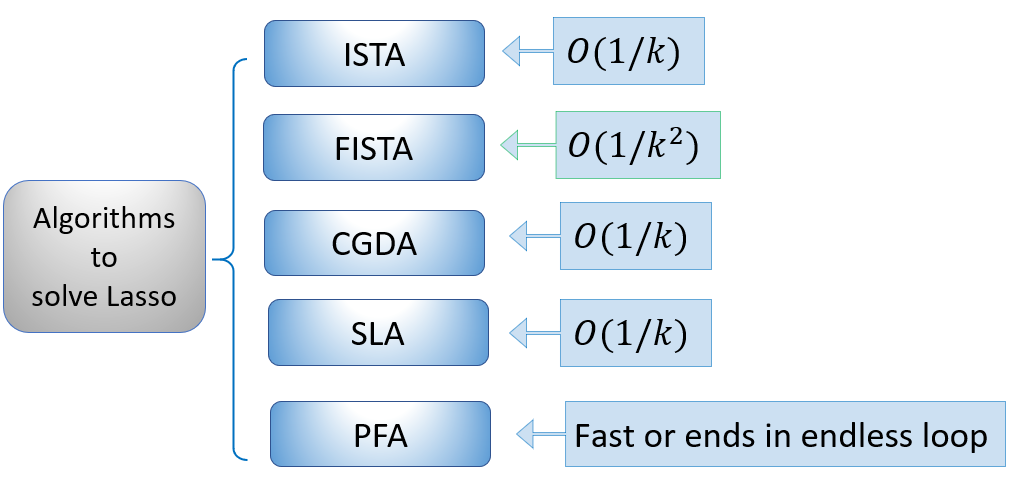}
\end{visAbstract}

%%%%%%%%%%%%%%%%%%%%%%%%%%%%%%%%%%%%%%%%%%%%%%%%%%%%%%%%%%%%
%%                 Introduction
%%%%%%%%%%%%%%%%%%%%%%%%%%%%%%%%%%%%%%%%%%%%%%%%%%%%%%%%%%%%

\section{Introduction}
\label{sec: introduction}

In the regression analysis, one has the data 
$$
  \mathcal D = \{y \in \mathbb R^n, X \in \mathbb R^{n\times p}\},
$$ 
where $y$ is the response vector and $X$ is the model matrix (of predictors). 
Here $n,p > 0$ refers to the number of observations and covariates, respectively. 
Given the above dataset $\mathcal D$, the linear regression model can be written as
$$
  y = X\beta^* + w,
$$
where $\beta^* \in \mathbb R^p$ is the ground truth of the regression coefficients desired to be estimated.
And the vector $w \in \mathbb R^n$ is the white-noise residual, i.e., $w_i \overset{i.i.d.}{\sim} N(0,\sigma^2)$ for any $i = 1, \ldots, n$.

When the number of covariates is greater than the number of observations, i.e., $p > n$, one prefers to select a subset of covariates and exclude the insignificant covariates. 
To realize the objective of variable selection, the least absolute shrinkage and selection operator (Lasso) \citep{tibshirani1996regression, santosa1986linear} can be used: 
\begin{equation}
	\label{equ: lasso estimator}
	\widehat\beta
	=
	\arg \min_\beta
	\left\{
	  F(\beta) := 
	  \frac{1}{2n}\|y-X\beta\|_2^2 + \lambda \|\beta\|_1
	\right\},
\end{equation}
where parameter $\lambda > 0$ controls the trade-off between the sparsity and model's goodness of fit.
The objective function in the Lasso method is $F(\beta)$, whose first term is a nice quadratic function and numerically amenable.
However, its second term of is not differentiable at the origin.
To minimize this objective function $F(\beta)$, there does not exist a closed-form minimizer and the existing algorithms are mostly iterative algorithms.

In this paper, we review representative algorithms to minimize $F(\beta)$ and compare their convergence rates. 
The convergence rate measures how quickly the sequence $\beta^{(0)}, \beta^{(1)}, \beta^{(2)}, \ldots$ approaches its optima $\widehat\beta$, where
$\beta^{(k)}$ is the iterative solution when minimizing $F(\beta)$ after $k$ iterations. 
And the convergence rate is commonly in terms of $k$ and the big $O$ notation, like $O(1/k), O(1/k^2)$.
In theory, an algorithm with convergence rate of $O(1/k^2)$ is more computationally efficient than that of $O(1/k)$.
Yet, it does not say anything on the average performance of the algorithm.
It is possible that an algorithm with convergence rate of $O(1/k)$ performs better in some cases than an algorithm with convergence rate of $O(1/k^2)$.

Please note that, we do not consider the selection of parameter $\lambda$ in this review, which by itself has a large literature.
Consequently, we don't include $\lambda$ in the notation $F(\beta)$.

In the remainder of this review, we first introduce some key preliminaries in Section \ref{sec: preliminary}.
Then we review five representative algorithms in Section \ref{sec: review of all algorithms}.
In Section \ref{sec: conclusion}, we give conclusions.

% ------------------------------------ %
%        preliminary                   %
% ------------------------------------ %
\section{Preliminaries}
\label{sec: preliminary}
In this section, we introduce some preliminaries, which helps readers to learn terminologies in statistical computations. 
The introduced terminologies include (1) Lipschitz continuous gradient, (2) convexity, (3) the (accelerate) gradient descent, (4) first/second order algorithms.

We begin with introducing two terms to describe functions, i.e., Lipschitz continuous gradient and convex functions.

\begin{mydef}[Lipschitz continuous gradient]
  A differentiable function $f(\cdot)$ has an Lipschitz continuous gradient $L$ if for some $L > 0$, one has 
  \begin{equation*}
	\left\|
	  \nabla f(x_1) - \nabla f(x_2)
	\right\|_2
	\leq 
	L\left\| x_1 - x_2 \right\|_2,
  \end{equation*}
  where $\nabla f(x)$ is the gradient of $f(x)$ at $x$.
\end{mydef}

\begin{figure}[htbp]
	\centering
	\begin{tabular}{cc}
		\includegraphics[width = 0.45\textwidth]{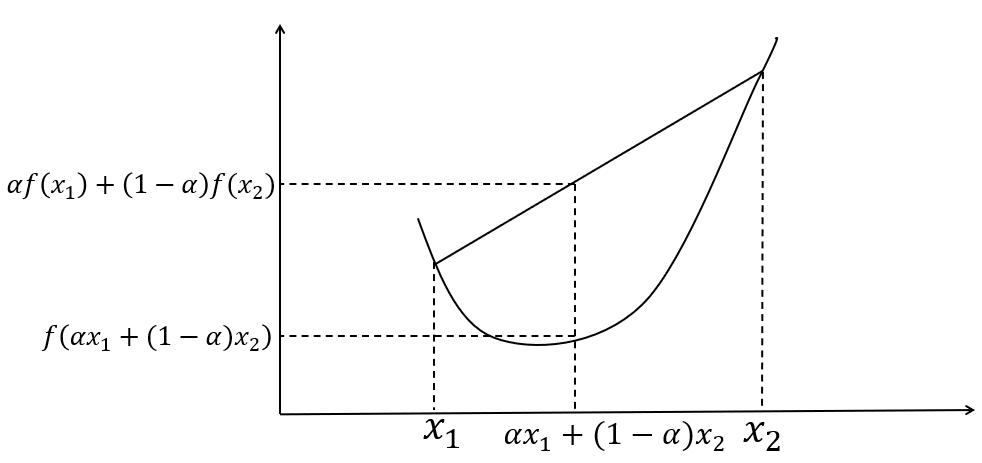} &
		\includegraphics[width = 0.45\textwidth]{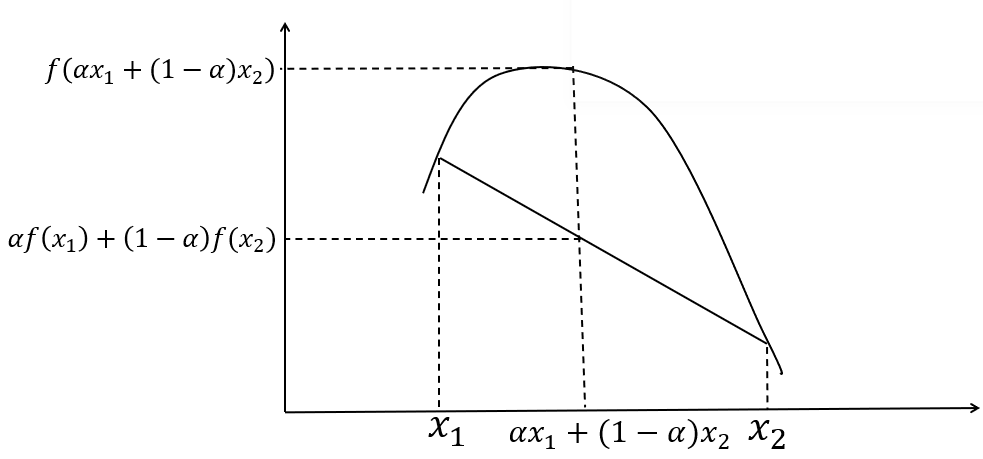}  \\
		(a) convex function &  (b) non-convex function  \\
	\end{tabular}
	\caption{Examples of convex and non-convex functions
	\label{fig: convex and non-convex} }
\end{figure}

\begin{mydef}[convex and strongly convex]
  A real-valued function $f(\cdot)$ is called convex, if the line segment between any two points on the graph of the function does not lie below the graph between the two points, i.e., $\forall x_1, x_2$ and $\alpha \in [0,1]$, one has
  $$
    f(\alpha x_1 + (1-\alpha) x_2)
    \leq
    \alpha f(x_1) + (1 - \alpha) f(x_2).
  $$
  If  there exist $\mu>0$, such that $\forall x_1, x_2$ we have 
  $$
    f(x_2) \geq f(x_1) + \nabla f(x_1)(x_2 - x_1) + \frac{\mu}{2} \left\| x_2 - x_1 \right\|_2^2,
  $$
  then we call $f(x)$ strongly convex.
\end{mydef}
For the above definition, one can verify that $f(x) = \log(x)$ is not a convex function, and $f(x) = x$ is convex but not strongly convex. 
On the other hand, $f(x) = x^2$ is convex and strongly convex. 
An example of a convex function is visualized in Figure \ref{fig: convex and non-convex}.
Besides, if a function both have Lipschitz continuous gradient as $L$ and is strongly convex with $\mu$, we call it $L$-smooth and $\mu$-strongly convex function.

Next, we present one version of gradient descent (GD) and accelerated gradient descent (AGD) algorithm, which build the foundations of our reviewed algorithms.
\begin{mydef}[GD algorithm]
\label{def: gd}
  Suppose one wants to minimize a convex and differentiable function $f: \mathbb R^p \to \mathbb R$, and that its gradient is Lipschitz continuous with constant $L$.
  To minimize $f(x)$ with respect to $x$, the gradient descent (GD) algorithm iterates as follows
  $$
    x^{(k)} = x^{(k-1)} - \gamma_{k-1} \nabla f(x^{(k-1)})
  $$
  for $k = 1, 2, \ldots$. 
  Here $x^{(k)}$ is the solution after $k$ itertions.
  And the hyper-parameter $\gamma_k > 0$ is the step size or the learning rate, which can be either fixed throughout the iterations, or decided by the backtracking line search.
\end{mydef}
In the sequence of the gradient descent iterations, one has a monotonic sequence $f(x^{(0)}) \geq f(x^{(1)}) \geq f(x^{(2)}) \geq \ldots$. 
Promisingly, the sequence $\{x^{(k)}\}_{k = 0,1, \ldots}$ converges to the desired local minimum.
The convergence rate of the GD algorithm is summarized in the following lemma.

\begin{mylemma}[GD convergence rate]
\label{lemma: gd convergence}
  Suppose the function $f: \mathbb R^p \to \mathbb R$ is convex and differentiable, and that its gradient is Lipschitz continuous with constant $L$.
  Then if one runs the GD algorithm in Definition \ref{def: gd} for $k$ iterations with a fixed step size $\gamma \leq 1/L$, it will yield a solution $f(x^{(k)})$ which satisfies
  $$
    f(x^{(k)}) - f(x^*) 
    \leq
    \frac{\left\| x^{(0)} - x^* \right\|_2^2}{2 \gamma k},
  $$
  where $f(x^*)$ is the optimal value.
  Otherwise, if one runs GD algorithm for $k$ iterations with step size $\gamma_k$ chosen by backtracking line search on each iteration $k$, it will yield a solution $f(x^{(k)})$ which satisfies
  $$
    f(x^{(k)}) - f(x^*) 
    \leq
    \frac{\left\| x^{(0)} - x^* \right\|_2^2}{2 \gamma_{\min} k},
  $$
  where $\gamma_{\min} = \min\{1, \beta/L\}$.
  Here $\beta \in (0, 1)$ is a hyper-parameter in the backtracking line search: in the $k$-th iteration, we start with $\gamma_k = 1$ and while 
  $
    f(x - \nabla f(x)) > f(x) - \frac{1}{2} \gamma \left\| \nabla f(x) \right\|_2^2,
  $
  we update $\gamma_k = \beta \gamma_k$.
\end{mylemma}

Intuitively, the above lemma indicates that the GD algorithm is guaranteed to converge and that it converges with rate $O(1/k)$.
Motivated by GD algorithm, researchers \citep{FISTAhistoryNesterov1983} proposed its accelerated version, called AGD algorithm.
It has been widely applied into many optimization problems to speed up the convergence rate \citep{AGDapply1, AGDapply2, FISTA, AGDapply4}.
Following please find one version of the AGD algorithm.

\begin{mydef}[AGD algorithm]
\label{def: agd}
  Suppose ones wants to  minimize a function $f = g + h: \mathbb R^p \to \mathbb R$, where $g$ is a convex and differentiable function and $h$ is a convex function. 
  To minimize $f(x)$ with respect to $x$, the accelerated gradient descent (AGD) algorithm iterates as follows
  \begin{eqnarray}
    y^{(k)}
    & = &
    \nonumber 
    x^{(k-1)} + \frac{k-2}{k+1} \left[ x^{(k-1)} - x^{(k-2)} \right]  \\
    x^{(k)}
    & = &
    prox_{t_k} 
    \left(
      y^{(k)} - \gamma_k \nabla g(y^{(k)})
    \right),
  \end{eqnarray}
  where
  $$
    \text{prox}_t(x)
    =
    \arg\min_{z \in \mathbb R^p} 
    \frac{1}{2t} 
    \left\| x - z\right\|_2^2 + h(z).
  $$
  Here $x^{(k)}$ is the solution after $k$ iterations. And $y^{(k)}$ is an auxiliary vector after $k$ iterations.
  The hyper-parameter $t_k$ is the step size or the learning rate, which can be either fixed throughout iterations, or decided by the backtracking line search.
\end{mydef}

Following the procedure of the AGD algorithm, we further introduce the convergence rate of the AGD algorithm \citep[Theorem 3.7]{lan2019lectures}.

\begin{mylemma}[AGD convergence rate]
\label{theo: appendix -- lan -- AGD -- convergence rate}
  Suppose a function $f = g + h: \mathbb R^p \to \mathbb R$ with $g$ as a convex and differentiable function and $h$ as a convex function. 
  Then if one runs the AGD algorithm in Definition \ref{def: agd} for $k$ iterations with a fixed step size $\gamma \leq 1/L$, it will yield a solution $f(x^{(k)})$ which satisfies
  $$
    f(x^{(k)}) - f(x^*) 
    \leq
    \frac{2 \left\| x^{(0)} - x^* \right\|_2^2}{\gamma (k+1)^2}.
  $$
  Otherwise, if one runs the AGD algorithm with a backtracking line search generated step size $\gamma_k \leq 1/L$, it will yield a solution $f(x^{(k)})$ which satisfies
  $$
    f(x^{(k)}) - f(x^*) 
    \leq
    \frac{2 \left\| x^{(0)} - x^* \right\|_2^2}{\gamma_{\min} (k+1)^2}.
  $$
  where $\gamma_{\min}$ is defined in Lemma \ref{lemma: gd convergence}.
\end{mylemma}

The above lemma indicates that, the AGD algorithm is guaranteed to converge and that it converges with rate $O(1/k^2)$. 
And compared with GD algorithm, the AGD algorithm has a faster convergence rate theoretically. 

Finally, we introduce the definition of \textit{first order algorithm} and \textit{second order algorithm}.

\begin{mydef}[first/second-order algorithm]
\label{def: first and second order method}
  Any algorithm that requires at most the gradient/first order derivative is a \textit{first order algorithm}.
  Any algorithm that uses any second order derivative is a \textit{second order algorithm}.
  The \textit{accelerated first order algorithm} is a particular type of algorithms that use multiple steps of gradients/first order derivatives. 
\end{mydef}

The well-known Newton–Raphson method is a second-order algorithm since it uses the second order derivatives (i.e., the Hessian).
The GD algorithm is a first order algorithm. 
The AGD algorithm is motivated by the first order algorithms to learn more ``history.''
Specifically, the classical first order algorithm uses the gradient at the immediate previous solution.
While the accelerated first order algorithms take advantage of the gradients at the previous two solutions, to learn from the ``history.''
Since it uses the historic information, it is also referred to as the \textit{momentum} algorithm.

\section{Review of Existing Algorithms to Solve Lasso}
\label{sec: review of all algorithms}
In this section, we presents five representative algorithms to solve Lasso: 
(1) iterative shrinkage threshold algorithm (ISTA), 
(2) fast iterative shrinkage-thresholding algorithms (FISTA), 
(3) coordinate gradient descent algorithm (CGDA), 
(4) smooth L1 algorithm (SLA) and 
(5) path following algorithm (PFA).

% ------------------------------------ %
%    ISTA method                       %
% ------------------------------------ %
    
\subsection{ISTA}
\label{sec: ista}

ISTA \citep{ISTA} is a first order method (see Defition \ref{def: first and second order method}), which targets on the minimization of a summation of two functions
$$
  f(\beta) + g(\beta),
$$
where $f(\beta): \mathbb R^p\rightarrow \mathbb R$ is smooth convex with a Lipschitz continuous gradient and $g(\beta): \mathbb R^p \rightarrow \mathbb R$ is continuous convex.
If we let 
$$
  f(\beta) = \frac{1}{2n} \left\| y - X \beta \right\|_2^2, 
  \;\;
  g(\beta) = \lambda\Vert\beta\Vert_1
$$ 
with $f(\beta)$'s Lipschitz continuous gradient $L$ taking the largest eigenvalue of matrix $X'X/n$,
then we can verify that the objective function in Lasso $F(\beta)$ is a special case of ISTA.

To minimize $F(\beta)$, at the $k$-th iteration, ISTA updates $\beta^{(k+1)}$ from $\beta^{(k)}$ by using the quadratic approximation function of $f(\beta)$ at value $\beta^{(k)}$:
\begin{equation}
	\label{equ: ISTA update1}
	\beta^{(k+1)} = \arg\min_\beta f(\beta^{(k)}) +\langle(\beta-\beta^{(k)}),\nabla f(\beta^{(k)})\rangle +\frac{\sigma_{\max}(X'X/n)}{2} \Vert\beta-\beta^{(k)}\Vert_2^2 +\lambda\Vert\beta\Vert_1.
\end{equation}
Here $\sigma_{\max}(X'X/n)$ is the maximal eigen-value of the matrix $X'X/n$.
Simple algebra shows that (ignoring constant terms in $\beta$), minimization of  \eqref{equ: ISTA update1} is equivalent to the minimization problem in the following equation:
\begin{equation}
	\label{equ: ISTA update2}
	\beta^{(k+1)}
	=
	\arg\min_\beta \;
	\frac{\sigma_{\max}(X'X/n)}{2}
	\left\Vert
	  \beta - 
	  \left(
	    \beta^{(k)} - \frac{\frac{1}{n}(X'X\beta^{(k)}- X'y)}{\sigma_{\max}(X'X/n)} \right)
	\right\Vert_2^2
	+\lambda\Vert\beta\Vert_1,
\end{equation}
where the soft-thresholding function in equation \eqref{equ: soft thresholding} can be used to solve the problem in equation \eqref{equ: ISTA update2}.
\begin{eqnarray}
	\label{equ: soft thresholding}
	S(x,\alpha)&=&\left\{
	\begin{array}{ll}
		x-\alpha, & \mbox{ if } x\geq\alpha, \\
		x+\alpha, & \mbox{ if } x\leq -\alpha, \\
		0, &   \mbox{ otherwise. }
	\end{array}
	\right.
\end{eqnarray}
Specifically, one can update $\beta^{(k+1)}$ from $\beta^{(k)}$ as
$$
  \beta^{(k+1)} 
  =
  S\left(
  \beta^{(k)} - \frac{1}{n\sigma_{\max}(X'X/n)}(X'X\beta^{(k)} - X'y), \lambda/\sigma_{\max}(X'X/n)
  \right).
$$
The summary of ISTA algorithm is presented in Algorithm \ref{alg: ISTA}.

\begin{algorithm}[H]
	\label{alg: ISTA}
	\caption{Iterative Shrinkage-Thresholding Algorithms (ISTA) }
	\LinesNumbered
	\KwIn{$y_{n\times1}, X_{n\times p}$, $L= \sigma_{\max}(X'X/n)$}
	\KwOut{$\beta^{(K)}$: an estimator of $\beta$ after $K$ iterations}
	\bfseries{initialization}: 	$\beta^{(0)}$ \\
	\For{$k = 0, 1, \ldots, K$}{
		$
		  \beta^{(k+1)} 
		  =
		  S(\beta^{(k)} - \frac{1}{nL}(X'X\beta^{(k)} - X'y), \lambda/L)
		$
	}
\end{algorithm}

In addition to the implementation of ISTA, we also develop the convergence analysis of ISTA in the following equation \citep[Theorem 3.1]{FISTA}.
To make it more clear, we list  \citep[Theorem 3.1]{FISTA} below with several changes of notation.
The notations are changed to be consistent with the terminology that are used in this paper.

\begin{mytheorem}
  Let $\left\{ \beta^{(k)}\right\}$ be the sequence generated by Algorithm \ref{alg: ISTA}.
  Then for any $k \geq 1$, we have
  \begin{equation}
	\label{equ: ISTA prediction error bound}
	F(\beta^{(k)})-F(\widehat\beta) \leq \frac{\sigma_{\max}(X'X/n)\Vert\beta^{(0)}-\widehat\beta\Vert_2^2}{2k}.
  \end{equation}
  Accordingly, the convergence rate of ISTA is $O(1/k)$.
\end{mytheorem}

From the above theorem and Algorithm \ref{alg: ISTA}, we find the computational complexity of ISTA is of order $O(k p^2)$, where $k$ is the number of iterations and $p$ is the number of coveriates. 
This is because that, line 3 in Algorithm \ref{alg: ISTA} shows the number of operations in one loop of ISTA is $O(p^2)$.
This is because that the main computation of each loop in ISTA is the matrix multiplication in $X'X \beta^{(k)}$.
Note that the matrix $X'X$ can be pre-calculated and saved, therefore, in one loop, ISTA's order of computational complexity is $p(2p-1)$ \citep{algebraForMatrixCalculation}.
Accordingly, the computational complexity of ISTA is $O(k p^2)$.

\subsection{FISTA}    

Motivated by ISTA, \cite{FISTA} developed an accelerated version FISTA.
The major difference of ISTA and FISTA is that ISTA is a first order method (see Definition \ref{def: first and second order method}), which only uses the gradient at the immediate previous solution.
On the other hand, FISTA is a accelerated first order algorithm method, which takes advantage of the gradients at previous two solutions, in order to learn from the history.

The updating rule of FISTA from $\beta^{(k)}$ to $\beta^{(k+1)}$ is articulated as follows.
FISTA first employs an auxiliary variable $\alpha^{(k)}$ in the second-order Taylor expansion step (i.e., the one in equation \eqref{equ: ISTA update1}):
\begin{equation}
	\label{equ: FISTA update1}
	\beta^{(k+1)} =
	\arg\min_\alpha f(\alpha^{(k)}) +
	\langle(\alpha-\alpha^{(k)}),\nabla f(\alpha^{(k)})\rangle +
	\frac{\sigma_{\max}(X'X/n)}{2} \Vert\alpha-\alpha^{(k)}\Vert_2^2 +
	\lambda\Vert\alpha\Vert_1,
\end{equation}
where $\alpha^{(k)}$ is a specific linear combination of the previous two estimator $\beta^{(k-1)}, \beta^{(k-2)}$.
In particular, we have 
$$
  \alpha^{(k)}
  =
  \beta^{(k-1)}
  +
  \frac{t_{k-1}-1}{t_{k}}
  \left[ \beta^{(k-1)}-\beta^{(k-2)} \right].
$$
In this way, FISTA constructs a ``momentum'' term $\beta^{(k-1)} - \beta^{(k-2)}$, and learns more historic information.
This idea improves the computational complexity, as you will see in the reminder of this section.
For completeness, we present FISTA in Algorithm \ref{alg: FISTA}.

\begin{algorithm}[H]
	\label{alg: FISTA}
	\caption{ Fast Iterative Shrinkage-Thresholding Algorithms (FISTA) }
	\LinesNumbered
	\KwIn{$y_{n\times1}, X_{n\times p}$, $L= \sigma_{\max}(X'X/n)$ }
	\KwOut{$\beta^{(K)}$: an estimator of $\beta$ after $K$ iterations}
	\bfseries{initialization}\;  	
	$\beta^{(0)}$ $\alpha^{(1)} = \beta^{(0)}$,  $t_1 = 1$ \\
	\For{ $k = 1, \ldots, K$ }{
		$\beta^{(k)}=S(\alpha^{(k)}-\frac{1}{nL}(X'X\alpha^{(k)}- X'y), \lambda/L)$
        \label{algLine: FISTA generation beta}   \\
		$t_{k+1}=\frac{1+\sqrt{1+4t_k^2}}{2}$\\
		$\alpha^{(k+1)}=\beta^{(k)}+\frac{t_k-1}{t_{k+1}}(\beta^{(k)}-\beta^{(k-1)})$
        \label{algLine: FISTA generation alpha}\\
	}
\end{algorithm}

FISTA has an improved convergence rate as $O(1/k^2)$, as compared $O(1/k)$ from ISTA \citep[Theorem 4.4]{FISTA}.
To make it more clear, we list \citep[Theorem 4.4]{FISTA} below with several changes of notation. 
The notations are changed to be consistent with the terminology that are used in this paper.

\begin{mytheorem}
  Let $\left\{ \beta^{(k)} \right\}$ be a sequence generated by Algorithm \ref{alg: FISTA}.
  Then for any $k \geq 1$, we have that
  \begin{equation}
	\label{equ: FISTA prediction error}
	F(\beta^{(k)})-F(\widehat\beta)
    \leq
    \frac{2\sigma_{\max}(X'X/n)\Vert\beta^{(0)}-\widehat\beta\Vert_2^2}{(k+1)^2}.
   \end{equation}
   Accordingly, the convergence rate of FISTA is $O(1/k^2)$.
\end{mytheorem}

By combining the conclusion in the above theorem with Algorithm \ref{alg: FISTA}, we find FISTA's computational complexity after $k$ iterations is $O(k p^2)$. 
This is because, the number of operations in one iteration of FISTA is still $O(p^2)$ (see line 4 in Algorithm \ref{alg: FISTA}).
Although for both ISTA and FISTA, they have the same number of operation in one loop, FISTA has improved convergence rate than ISTA.
Specifically, after running both $k$ iterations, FISTA's output is more closer to the optima than that from ISTA, given its faster convergence rate.
   
% ------------------------------------ %
%        literature                    %
%    coordinate descent method         %
% ------------------------------------ %
\subsection{CGDA}
The aforementioned two algorithms (ISTA and FISTA) update their estimates globally in one iteration loop. 
The algorithm introduced in this section, CGDA, updates the estimates one coordinate at a time. 
Specifically it cyclically chooses one coordinate at a time and performs a simple analytical update.
Such an approach is called \textit{coordinate gradient descent}.
This approach has been proposed for the Lasso problem for a number of times, but only after \cite{glmnet}, was its power fully appreciated.
Early research work on the coordinate descent include the discovery by \cite{CDhistoryHildreth1957} and \cite{CDhistoryWarga1963}, and the convergence analysis by \cite{CDhistoryTseng2001}.
There are research work done on the applications of coordinate descent on Lasso problems, such as \cite{CDhistoryFu1998}, \cite{CDhistoryShevade2003}, \cite{CDhistoryFriedman2007}, \cite{CDhistoryWu2008}, and so on.

CGDA is widely used and the corresponding R package is named \textit{glmnet} \citep{glmnet}.
In CGDA, the updating rule from $\beta^{(k)}$ to $\beta^{(k+1)}$ is that, it optimizes with respect to only the $j$-th entry of $\beta^{(k+1)}$ for a selected $j=1,\cdots, p$.
And the gradient at $\beta_j^{(k)}$ is used for the updating process:
\begin{equation}
\label{equ: CD gradient information}
  \frac{\partial}{\partial \beta_j} F(\beta^{(k)})
  =
  \frac{1}{n}
  \left( e_j' X'X \beta^{(k)}  - y'X e_j\right)
  +
  \lambda \; \text{sign}(\beta_j)
\end{equation}
where $e_j$ is a vector of length $p$, whose entries are all zero expect that the $j$-th entry is equal to $1$.
Imposing the gradient in \eqref{equ: CD gradient information} to be $0$,  we can solve for $\beta^{(k+1)}_j$ as follows:
$$
\beta^{(k+1)}_j
=
S\left(
y' X e_j - \sum_{l \neq j} \left( X'X \right)_{jl} \beta^{(k)}_k, n\lambda
\right)
\bigg/ \left( X'X\right)_{jj},
$$
where $S(\cdot)$ is the soft-thresholding function defined in \eqref{equ: soft thresholding} and $(X'X)_{ij}$ is the $(i,j)$-th entry of the matrix $X'X$.
The above implementation is summarized in Algorithm \ref{alg: CD}.

\begin{algorithm}[H]
	\label{alg: CD}
	\caption{Coordinate Gradient Descent Algorithm (CGDA) }
	\LinesNumbered
	\KwIn{$y_{n\times1}, X_{n\times p}$, $\lambda$ }
	\KwOut{$\beta^{(K)}$: an estimator of $\beta$ after $K$ iterations}
	\bfseries{initialization:} $\beta^{(0)}$\\
	\For{$k = 0, 1, \ldots, K$ }{
		\For {$j=1\cdots p$}{
			$
			   \beta^{(k+1)}_j
               =
               S\left( y' X e_j - \sum_{l \neq j} \left( X'X \right)_{jl} \beta^{(k)}_k, n\lambda \right)
               \bigg/
               \left( X'X\right)_{jj} 
            $
               \label{algLine: CD generation}
		}
	}
\end{algorithm}

As reflected by Corollary 3.8 in \cite{CDconvergence}, we find the convergence rate of CGDA is $O(1/k)$
Here we list this corollary as a theorem below.
We changed several notations to adopt the terminology in this paper.

\begin{mytheorem}
  Let $\left\{ \beta^{(k)}  \right\}$ be the sequence generated by in Algorithm \ref{alg: CD}.
  Then we have that
  \begin{equation}
	\label{equ: CD prediction error}
	F(\beta^{(k)})-F(\widehat\beta)
    \leq
    \frac{4\sigma_{\max}(X'X/n)(1+p)\Vert\beta^{(0)}-\widehat\beta\Vert_2^2}{k+(8/p)}.
  \end{equation}
  Accordingly, the convergence rate of CGDA is $O(1/k)$.
\end{mytheorem}

Compared with ISTA and FISTA, we find CGDA share the same order of convergence rate as ISTA, which is less efficient than FISTA. 
This is because that, both ISTA and CGDA are first order algorithm (see Definition \ref{def: first and second order method}), while the FISTA is the accelerated first order algorithm.
For the first order algorithm, it utilize the previous gradient, while the accelerated first order algorithm takes advantage of the previous two graidents and learns more history from the previous two gradients. 
This momentum-learning ability allows FISTA to have faster convergence rate.

After reviewing the algorithm of CGDA, we develop its computational complexity.
Firstly, the number of operations in each loop of CGDA is $O(p^2)$.
It can be explained by the following two reasons.
(i) While updating $\beta^{(k+1)}_j$ (line 4 in Algorithm \ref{alg: CD}), it costs $O(p)$ operations because of
$
\sum_{l \neq j} \left( X'X \right)_{jl} \beta^{(k)}_k
$.
(ii) From line 3 in Algorithm \ref{alg: CD}, we can see that all $p$ entries of $\beta^{(k+1)}$ are updated one by one.
Combining (i) and (ii), we can see that the number of operations need in one loop of CD is of the order $O(p^2)$.
Therefore, CGDA's computational complexity after $k$ iterations is $O(k p^2)$.

\subsection{SLA}
\label{sec: sla}

The aforementioned three methods (ISTA, FISTA, CGDA) targets directly at the minimization of $F(\beta)$.
Different from them, SLA \citep{smoothlassoClass01} aims to minimize a smooth surrogate of $F(\beta)$.
And the surrogate commonly happens to the $\ell_1$ penalty term.
Recall the $\ell_1$ penalty is essentially the absolute function:
\begin{equation}
\label{equ: x to 2 max}
    |x| = \max\{x, 0\} + \max\{-x, 0\}, \;\; \forall x \in \mathbb R.
\end{equation}
And without much pain, one can find the non-differentiability of $|x|$ at the origin makes it hard to enable fast convergence rate when applying the gradient descent method. 

\begin{figure}[htbp]
    \centering
    \includegraphics[width = 0.5\textwidth]{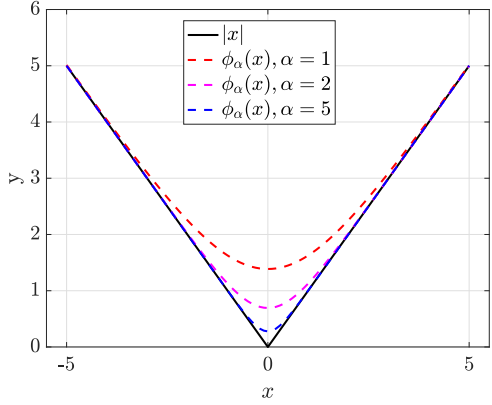}
    \caption{The closeness between $\phi_{\alpha}(x)$ and $|x|$ under different value of $\alpha$
    \label{fig: smooth L1 approx}}
\end{figure}

To get rid of non-differentiability of $|x|$, \cite{smoothlassoClass01} proposed one of its surrogate function:
\begin{eqnarray}
    |x|
    & \approx & \label{equ: abs x to max to sigmoid}
    \left[ x + \frac{1}{\alpha} \log\left( 1 + \exp(-\alpha x)\right) \right]
    +
    \left[ -x + \frac{1}{\alpha} \log\left( 1 + \exp(\alpha x)\right) \right] \\
    & = & \nonumber
    \underbrace{
    \frac{1}{\alpha}
    \left[
      \log\left( 1 + \exp(-\alpha x)\right)
      +
      \log\left(1 + \exp(\alpha x)\right)
    \right]
    }_{\phi_{\alpha}(x)}
\end{eqnarray}
Here $\alpha$ is a hyper-parameter controlling the closeness between $|x|$ and its surrogates $\phi_{\alpha}(x)$.
The curve of this surrogate function $\phi_{\alpha}(x)$ is available in Figure \ref{fig: smooth L1 approx}.
From this figure, one can tell that a large value of $\alpha$ makes better approximation to $|x|$. 
As suggested by \cite{smoothlassoClass01}, to select an appropriate $\alpha$, one can always begins from a small $\alpha$ where the quadratic approximation from $\phi_{\alpha}(x)$ is appropriate, and terminates at a sufficiently large value of $\alpha$.

The above surrogate function is motivated by the non-negative projection operator $\max\{x, 0\}$ in \eqref{equ: x to 2 max}, which can be smoothly approximated by the integral of a sigmoid function as shown in \eqref{equ: abs x to max to sigmoid}.

A nice feature of the surrogate function in \eqref{equ: abs x to max to sigmoid} is its twice differentiability.
Consequently, the AGD algorithm (see Definition \ref{def: agd}) is applicable to the surrogate function:
$$
  F_\alpha(\beta)
  =
  \frac{1}{2n}\|y-X\beta\|_2^2 + \lambda \sum_{i=1}^{p}\phi_\alpha(\beta_i),
$$
where 
$
  \phi_\alpha(\beta_i) 
  = 
  \frac{1}{\alpha}
  \left[
    \log\left( 1 + \exp(-\alpha \beta_i)\right)
    +
    \log\left( 1 + \exp(\alpha \beta_i)\right)
  \right]
$
for any $i = 1, \ldots, p$.
The pseudo code of SLA algorithm is displayed in Algorithm \ref{alg: SL}.

\begin{algorithm}[H]
	\label{alg: SL}
	\caption{Smooth L1 algorithm (SLA) }
	\LinesNumbered
	\KwIn{$y_{n\times1}, X_{n\times p}$, $\mu=\left[\sigma^2_{\max}(X/\sqrt n)+\lambda\alpha/2\right]^{-1}$}
	\KwOut{$\beta^{(K)}$: an estimator of $\beta$ after $K$ iterations}
	\bfseries{initialization: $\beta^{(0)}$} \\
	\For{ $k = 1, \ldots, K$  }{  	
		$w^{(k+1)}=\beta^{(k)}+\frac{k-2}{k+1}(\beta^{(k)}-\beta^{(k-1)})$\\
		$\beta^{(k+1)}=w^{(k+1)}-\mu\nabla F_\alpha( w^{(k)})$
        \label{algLine: SL generation}\\
	}
\end{algorithm}

As proved by \cite{smoothlasso}, the approximation error of $\beta^{(k)}$ in SLA is shown in equation \eqref{equ: SL prediction error}.
\begin{mytheorem}
  Let $\left\{ \beta^{(k)}\right\}$ be a sequence generated by Algorithm \ref{alg: SL}.
  Then we have
  \begin{equation}
	\label{equ: SL prediction error}
	F(\beta^{(k)})-F(\widehat\beta)
    \leq  
    \frac{4\left\|\beta^{(0)}-\widehat\beta\right\|_2^2\sigma_{\max}^2(\frac{X}{\sqrt{n}})}{k^2}
    +
    \frac{4\sqrt{2\lambda n\log2}\left\|\beta^{(0)}-\widehat\beta\right\|_2}{k}.
  \end{equation}
  Accordingly, the convergence rate of SLA is $O(1/k)$.
\end{mytheorem}

From the above theorem, we find the convergence rate of SLA is decided by the summation of two terms. 
The first term origins from the convergence rate of AGD (recall Lemma \ref{theo: appendix -- lan -- AGD -- convergence rate}). 
And the second term is caused by the difference between $F_{\alpha}(\beta)$ and $F(\beta)$: if one aims at the minimization of $F_{\alpha}(\beta)$, then the convergence rate is $O(1/k^2)$; however, if one aims at the minimization of $F(\beta)$, then the convergence rate is slowed by the difference between $F(\beta)$ and $F_{\alpha}(\beta)$.

For SLA's computational complexity, it mainly lies in the calculation of $\nabla F_\alpha(w)=\frac{X'X}{n}w-\frac{X'y}{n}+v$, where the $v$ is a vector of length $p$, whose $i$-th entry is $\frac{-2}{w_i^2}\log(1+e^{\alpha w_i})+\frac{2\alpha e^{\alpha w_i}}{w_i(1+e^{\alpha w_i})}-1$.
Accordingly, the main computational effort of one loop of SLA is the matrix multiplication in $X'X w^{(k)}$, which cost $O(p^2)$ operations.
Thus, SLA's computational complexity after $k$ iterations is of order $O(k p^2)$.

\subsection{PFA}

Another representative algorithm to minimize $F(\beta)$ utilizes the path following idea \citep{park2007l1, rosset2007piecewise,  tibshirani2011solution}, and we call this type of algorithms as PFA.
As the name suggests, PFA forms a path of the penalty parameter $\lambda_0, \lambda_1, \ldots, \lambda_K$.
Accordingly, it gets a sequence of $\beta$ estimated under this sequence of $\lambda$.
And we denote this sequence of $\widehat\beta$ as $\widehat\beta(\lambda_0), \widehat\beta(\lambda_1), \ldots, \widehat\beta(\lambda_K)$.

The first key block of PFA is to identify the sequence of the the penalty parameter, i.e., $\lambda_0, \lambda_1, \ldots, \lambda_K$. 
Before introducing the identification of $\lambda$ sequence, we introduce a terminology called \textit{support set}, which is useful in the identification.
\begin{mydef}[support set]
  For any $\beta \in \mathbb R^p$, its support set is the collection of indexes, whose entries are non-zero:
  $$
    S(\beta)
    = 
    \left\{ 
      i: \beta_i \neq 0
    \right\}.
  $$
  Here $\beta_i$ is the $i$-th entry of $\beta$. 
  And $|S(\beta)|$ measures the number of elements in the set $S(\beta)$. 
\end{mydef}
To identify the $\lambda$ sequence, PFA begins with a large $\lambda_0$, which makes the estimated $\widehat\beta(\lambda_0) = 0$, and accordingly its support set is an empty set, i.e., $S(\widehat\beta(\lambda_0)) = \emptyset$.
Then it tries to identify a sequence of the penalty parameter $\lambda$ as follows:
$$
  \lambda_0
  > \lambda_1
  > \lambda_2
  > \ldots
  > \lambda_{K-1}
  > \lambda_K = 0,
$$
such that for any $k \geq 1$, when we have
$
  \lambda \in [\lambda_k, \lambda_{k-1}],
$
the support set of $\widehat\beta(\lambda)$ (which is a function of $\lambda$) i.e., $S_k$, remains unchanged.
Moreover, within the interval $[\lambda_k, \lambda_{k-1}]$, vector $\widehat\beta(\lambda)$ elementwisely is a linear function of $\lambda$.
However, when one is over the kink point, the support is changed/enlarged, i.e., we have $S_{k} \neq S_{k-1}$ or even $S_{k} \subseteq S_{k-1}$.

The second key block of PFA is the estimation of $\widehat\beta(\lambda_k)$ given $\lambda_0, \ldots, \lambda_k$.
Instead of estimating $\beta$ directly under $\lambda_k$, PFA takes advantage of the correlation between $\widehat\beta(\lambda_k)$ and $\widehat\beta(\lambda_{k-1})$.
In the reminder of this section, we show this correlation and a concrete example is available in Section \ref{discussion -- path following -- drawback1}.
For a general solution derived by PFA, we know $\widehat\beta(\lambda)$ is the minimizer of \eqref{equ: lasso estimator}.
So it must satisfy the first order condition of \eqref{equ: lasso estimator}:
\begin{equation}
\label{equ: first-order condition of lasso in path following}
  q - \lambda \text{sign}( \widehat\beta(\lambda)) = X'X \widehat\beta(\lambda),
\end{equation}
where $q = X'y$ and $\text{sign}( \widehat\beta(\lambda))$ is a vector, whose $i$-th component is the sign function of $\widehat\beta(\lambda)$:
$$
  \text{sign}(\widehat\beta_i(\lambda)) =
  \left\{
  \begin{array}{cc}
    1                  & \text{if } \widehat\beta_i(\lambda)>0 \\
    -1                 & \text{if } \widehat\beta_i(\lambda)<0 \\
    \left[-1,1 \right] & \text{if } \widehat\beta_i(\lambda) = 0
  \end{array}
  \right..
$$
If we divide the indices of $q, \beta, X$ into
$
  S= \{i: \widehat\beta_i(\lambda) \neq 0, \;\forall\; i= 1,\ldots,p\}
$
and its complements $S^c$, then we can rewrite \eqref{equ: first-order condition of lasso in path following} as
\begin{equation*}
  \left(
  \begin{array}{c}
    q_S  \\
    q_{S^c}
  \end{array}
  \right)
  -
  \left(
  \begin{array}{c}
    \lambda \text{sign}( \widehat\beta_S(\lambda)) \\
    \lambda \text{sign}( \widehat\beta_{S^c}(\lambda))
  \end{array}
  \right)
  =
  \left(
  \begin{array}{cc}
    X_S' X_S & X_S' X_{S^c} \\
    X_{S^c}' X_S & X_{S^c}' X_{S^c}
  \end{array}
  \right)
  \left(
  \begin{array}{c}
    \widehat\beta_S(\lambda) \\
    0
  \end{array}
  \right),
\end{equation*}
where $\widehat\beta_S(\lambda)$ is the subvector of $\beta$ only contains elements whose indices are in $S$ and $\widehat\beta_{S^c}(\lambda)$ is the complement of $\beta_S$.
Besides, $\text{sign}(\widehat\beta_S(\lambda))$ is the subset of $\text{sign}(\widehat\beta(\lambda))$, only contains the elements whose indices are in $S$, and $\text{sign}(\widehat\beta_{S^c}(\lambda))$ is the complement to $\text{sign}(\widehat\beta_S(\lambda))$.
Matrix $X_S$ is the columns of $X$ whose indices are in $S$, and $X_{S^c}$ is the complement of $X_S$.

Suppose we are interested in parameter estimated under $\lambda$ and $\lambda - \Delta$, i.e., $\widehat\beta(\lambda), \widehat\beta(\lambda - \Delta)$, for any $\Delta\in(0,\lambda)$.
Then $\widehat\beta(\lambda), \widehat\beta(\lambda - \Delta)$ must satisfy the following two system of equations:
\begin{equation}
\label{equ: lambda}
  \left\{
  \begin{array}{rcl}
    q_S - \lambda \text{sign}( \widehat\beta_S(\lambda))
    &=&
    X_S' X_S \widehat\beta_S(\lambda) \\
    q_{S^c} - \lambda \text{sign}(\widehat\beta_{S^c}(\lambda) )
    &=&
    X_{S^c}' X_S \widehat\beta_S(\lambda)
  \end{array}
  \right.,
\end{equation}

\begin{equation}
\label{equ: lambda - delta}
  \left\{
  \begin{array}{rcl}
    q_S - (\lambda - \Delta) \text{sign}(\widehat\beta_S(\lambda - \Delta))
    & = &
    X_S' X_S \widehat\beta_S(\lambda - \Delta) \\
    q_{S^c} - (\lambda - \Delta) \text{sign}(\widehat\beta_{S^c}(\lambda - \Delta) )
    & = &
    X_{S^c}' X_S \widehat\beta_S(\lambda - \Delta)
  \end{array}
  \right..
\end{equation}
From the above two system of equations, we have the following:
\begin{equation}
\label{equ: lambda -> lambda-Delta}
   - (\lambda - \Delta) \text{sign}(\widehat\beta_{S^c} (\lambda - \Delta))
  =
   - \lambda \text{sign}(\widehat\beta_{S^c}(\lambda)) + \Delta X_{S^c}' X_S (X_S' X_S)^{-1} \text{sign}(\widehat\beta_S (\lambda)).
\end{equation}
That is, if one decreases $\lambda$ to $\lambda-\Delta$, one must strictly follow \eqref{equ: lambda -> lambda-Delta}.
The above equation is useful to find the support set of $\widehat\beta(\lambda - \Delta)$. 
After the support set is available, one can use the linear regression method, restricted to the support set, to get the estimation of $\widehat\beta(\lambda - \Delta)$.

$$
  - (\lambda - \Delta) \text{sign}(\widehat\beta_{S^c} (\lambda - \Delta))
  =
  - \lambda \text{sign}(\widehat\beta_{S^c}(\lambda)) + \Delta X_{S^c}' X_S (X_S' X_S)^{-1} \text{sign}(\widehat\beta_S (\lambda)).
$$
The pseudo code of the path-following algorithm is summarized as in Algorithm \ref{alg: path following}.

\begin{algorithm}[H]
	\label{alg: path following}
	\caption{Path following algorithm (PFA)}
	\LinesNumbered
	\KwIn{$y_{n\times1}, X_{n\times p}, \lambda$}
	\KwOut{an estimator of $\beta$ under penalty parameter $\lambda$}
	\bfseries{initialization:} 	$\lambda^{(0)}, k = 0$\\
	\While{ $\lambda^{(k)} > \lambda$  }{  	
		$
		  \lambda_{k+1} =
		  \sup
		  \left\{
		    \lambda:
		    \begin{array}{l}
		    \lambda < \lambda_k \textnormal{ and } \\
		    \forall \lambda', \lambda'' \in (\lambda, \lambda_k), \; S(\lambda') = S(\lambda'') \textnormal{ and } \\
		    \forall \lambda' \in [0, \lambda), \; 
		    S(\lambda') \neq S(\lambda)
		    \end{array}
		 \right\}
	    $\\
		$
		  \widehat\beta (\lambda_{k+1}) = 
		  \arg \min_\beta
		  \left\{
		    \frac{1}{2n}\|y - X\beta\|_2^2 + \lambda_{k+1} \|\beta\|_1
		  \right\}
		$\\
		$k=k+1$ 
	}
\end{algorithm}

For the computational effort, it mainly decided by the length of the $\lambda$ sequence, i.e., $K$.
If $K$ is small, then PFA is efficient: it only requires $O(n K p^2)$ numerical operations.
Compared with ISTA, FISTA, CDGA and SLA, when their number of iterations $k$ is larger than $nK$, then PFA is more computationally efficient theoretically. 
In particular, if the size of supports are strictly increasing, i.e., we have
$$
  |S_{k-1}| < |S_k| \;\; \forall k \geq 1,
$$
then we have $K \leq p$, and accordingly the total number of  numerical operations of PFA can be bounded by $O(n p^3)$.
Under this scenario, PFA is theoretically faster than ISTA, FISTA, CSDA and SLA, if they iterate more than $np$ iterations.

Yet, the contemporary literature indicates that the upper bound of $K$ is an an open question \citep{tibshirani2011solution, rosset2007piecewise}.
With unknown $K$, it is not theoretically guaranteed PFA converges. 
And it is possible that its convergence rate is low, considering that the maximum number of $K$ can be as large as $2^p$.

In additional to the unpredictable convergence rate, PFA has another limitation: it doesn't work for general cases. 
The current literature only establishes PFA in special situations.
Yet, it might fail to deliver the solution under some cases. 
In Section \ref{discussion -- path following -- drawback1}, we give a counter example where PFA fails.

\subsubsection{A Counter Example where PFA Fails}
\label{discussion -- path following -- drawback1}

In this section, we offer a concrete counter example where PFA fails.
This counter example represents a general category of design matrix $X$ and coefficient $\beta$.
And we use the following counter example to argue that PFA does not work in the most general setting.

The counter example is designed as follows.
Suppose
$$
  \beta_1 > \beta_2 > \beta_3 > \beta_4 = \beta_5 = \ldots = \beta_p = 0.
$$
The model matrix
$
  X = (X_1, X_2, X_3,\ldots, X_p)
$,
where
$
  X_1, X_2 \in \mathbb R^n
$
is the first two columns from a orthogonal matrix
$
  (X_1, X_2, \widetilde X_3, \ldots, \widetilde X_p)
$,
and for $j \geq 3$, we have
$
  X_j = \alpha_j X_1 + (1-\alpha_j) X_2 + \sqrt{1 - \alpha_j^2 - (1-\alpha_j)^2} \widetilde X_j$ with $\alpha_j \in (0,1)
$.
The response vector $y$ is generated by
$$
  y = \sum_{j=1}^{p} \beta_j X_j.
$$
If $\beta_1, \beta_2$ are very large number, say, 200, 100, and $\beta_3$ is not that large, say, 1.
Then PFA works as follows:
\begin{itemize}
  \item Loop 0: We start with $\lambda_0 = +\infty$, then we know that $\widehat\beta(\lambda_0) = 0$ and $S_0 = \emptyset$.
  \item Loop 1: When $\lambda$ changes from $\lambda_0 = +\infty$ to $\lambda_1 = \| q \|_\infty$, from \eqref{equ: first-order condition of lasso in path following}, we know that $S_1 = \{ 1 \}$.
  \item Loop 2:  Similar to the first loop, when $\lambda$ decrease to $\lambda_2$, we have $S_2 = \{1, 2 \}$.
  \item Loop 3:  This is where problem happens.
  From \eqref{equ: lambda -> lambda-Delta}, we know that $\forall \lambda_2 - \Delta \in (\lambda_3, \lambda_2]$, we have
  $$
    \text{sign}(\widehat\beta_{S_2^c} (\lambda - \Delta))
    =
    X_{S_2^c}' X_{S_2} (X_{S_2}' X_{S_2})^{-1} \text{sign}(\widehat\beta_{S_2}(\lambda)).
  $$
  Since
  $
    \text{sign}(\widehat\beta_{S_2}(\lambda)) = (1, 1)'
  $
  and
  $
    X_{S_2} = (X_1, X_2), X_{S_2^c} = (X_3, \ldots, X_p)
  $,
  we have the right hand side of the above equation as a all-one vector, i.e, $(1, 1, \ldots, 1)'$.
  To make the left hand side $\text{sign}(\widehat\beta_{S_2^c} (\lambda_2 - \Delta))$ equals to $(1, 1, \ldots, 1)'$, we can only take $\Delta = \lambda_2$, which gives us $S_3 = \{1,2,3,\ldots, p\}$.
\end{itemize}
However, from the data generalization, we know that the true support set is $\{1,2,3\}$.
Therefore, one will not be able to develop a PFA to realize correct support set recovery.
At least not in the sense of inserting one at a time to the support set. 
In the above example, since a path following approach can only visit three possible subsets, it won't solve the Lasso in general.

\subsection{LARS}
\label{sec: lars}
In the statistical community, there has been an algorithm universally used in the last decades.
It is called LARS, which can be regarded as an advanced forward selection method. 
It is originally developed by \cite{efron2004least} and later a R package named \textit{lars} full filled its implementations.
Nowadays, the LARS has decreasing popularity given its limitation in computation efficiency. 
So we will briefly introduce this algorithm in this review.

The main idea of LARS is articulated as follows.
We start with all coefficients equal to zero and find the predictor most correlated with the response, say $x_{j_1}$. 
We take the largest step possible in the direction of this predictor until some other predictor, say $x_{j_2}$, has as much correlation with the current residual. 
At this point LARS parts company with forward selection. 
Instead of continuing along $x_{j_1}$, LARS proceeds in a direction
equiangular between the two predictors until a third variable $x_{j_3}$ earns its way into the ``most correlated'' set. 
LARS then proceeds equiangularly between $x_{j_1}$, $x_{j_2}$ and $x_{j_3}$, that is, along the ``least angle direction,'' until a fourth variable enters, and so on.

Its detailed implementation is listed as follows \citep{tibshirani2009simple}.

\begin{itemize}
    \item Step 1: start with all coefficients $\beta$  equal to zero.
    \item Step 2: find the predictor $x_j$ most correlated with $y$.
    \item Step 3: increase the coefficient $\beta_j$ in the direction of the sign of its correlation with $y$. Take residuals $r=y-\hat y$ along the way. Stop when some other predictor $x_{k}$ has as much correlation with $r$ as $x_j$ has.
    \item Step 4: increase $(\beta_{j}, \beta_{k})$ in their joint least squares direction, until some other predictor $x_m$ has as much correlation with the residual $r$.
    \item Step 5: increase $(\beta _{j}, \beta_{k}, \beta_m)$ in their joint least squares direction, until some other predictor $x_n$ has as much correlation with the residual $r$.
    \item $\dots$
\end{itemize}
The above procedure continues until all predictors are in the model.
However, to our best knowledge, it is not clear when LARS stops, viewing from the existing literature. 
The LARS procedure may take many steps before it stops (see examples in \cite{turlach2004least}.). 
And to be worse, LARS is not workable for all cases \citep{turlach2004least}. 
Given this, the LARS algorithm is not as frequently used as the other algorithms reviewed in Section \ref{sec: ista} - \ref{sec: sla}.

\section{Conclusions}
\label{sec: conclusion}

Lasso is a regression method, which can realize both variable selection and regularization.
When one aims at the minimization of the objective function in Lasso, the $\ell_1$ penalty raises the computational challenge due to its non-differentiability.
To overcome this challenge, various iterative algorithms are proposed, including first order algorithm (like ISTA, CSDA, SLA) and accelerated first order algorithm (like FISTA).
Additionally, the path following idea is utilized to solve Lasso (like PFA). 
Comparing the convergence rate of the five algorithms, we find FISTA gives the best and relatively robust performance. 
Specifically, FISTA's convergence rate is $O(1/k^2)$, while the convergence rate of ISTA, CGDA and SLA are all $O(1/k)$.
For PFA, its might be faster than FISTA under some special cases.
However, there is no theoretical guarantee that it is faster than FISTA under general cases (see a counter example in Section \ref{discussion -- path following -- drawback1}).
The above comparison is summarized in Table \ref{table: pros and cons} and hope this comprehensive summary helps readers to learn more details about optimization with $\ell_1$ penalty and to facilitate their future research.

\begin{table}[htbp]
  \caption{Pros and cons of the five reviewed algorithm to solve Lasso 
  \label{table: pros and cons}}
  \centering
  \begin{adjustbox}{max width=1\textwidth}
  \begin{threeparttable}
  \begin{tabular}{p{0.1\textwidth}|p{0.46\textwidth} | p{0.44\textwidth}}
    \hline
    & Pros & Cons \\
    \hline
    ISTA & 
    \pbox{20cm}{ 
    Convergence rate is $O(1/k)$. 
    }
    & 
    \pbox{20cm}{ 
    Computationally inefficient than  \\ 
    FISTA or PFA under special cases.}
    
    \\
    \hline
    % ------------- %
    FISTA & 
    \pbox{20cm}{ 
    Convergence rate is $O(1/k^2)$ which is \\
    faster than ISTA, CGDA and SLA. \\ 
    }
    & 
    \pbox{20cm}{ 
    Could be computationally inefficient \\
    than PFA under special cases.}
    \\
    \hline
    % ------------- %
    CGDA & 
    \pbox{20cm}{ 
    Convergence rate is $O(1/k)$.
    }
    &
    \pbox{20cm}{ 
    Computationally inefficient than  \\ 
    FISTA or PFA under special cases.}
    
    \\
     \hline
    % ------------- %
    SLA &
    \pbox{20cm}{ 
    Convergence rate is $O(1/k)$.
    }
    &
    \pbox{20cm}{ 
    (1) Computationally inefficient than  \\ 
    FISTA or PFA under special cases;\\
    (2) The output is not the exactly $\widehat\beta$\\
    desired in Lasso.}
    
    \\
     \hline
    % ------------- %
    Path-following & 
    \pbox{20cm}{ 
    (1) Could be computationally more  \\
    efficient than FISTA under special cases. \\
    (2) Once the solution path is available, \\
    one can get the estimation of $\beta$ \\
    under any value of $\lambda$.
    }
    &
    \pbox{20cm}{ 
    (1) Not workable for generalized cases; \\
    (2) Computation complexity can be \\
    unbounded.
    }
    \\
    
	\hline
  \end{tabular}
  %\begin{tablenotes}
  %\footnotesize
    %\item[1] The computational complexity is defined in Definition \ref{def: epsilon precision}.
  %\end{tablenotes}
  \end{threeparttable}
  \end{adjustbox}
\end{table}

\section*{Funding Information}

This project is partially supported by the Transdisciplinary Research Institute for Advancing Data Science (TRIAD), http://triad.gatech.edu, which is a part of the TRIPODS program at NSF and locates at Georgia Tech, enabled by the NSF grant CCF-1740776. The authors are also partially sponsored by NSF grants 1613152 and 2015363.

\section*{Acknowledgments}

The authors would like to thank the editors of WIREs Computational Statistics for reviewing our proposal, and the editorial office for tracking our paper submission status.

\bibliographystyle{apalike}
\bibliography{reference}

\end{document}